\documentclass[a4paper,dvips,12pt]{article}
\usepackage{epsf}
\usepackage[dvips]{graphicx}
\usepackage{tabularx}
\usepackage{latexsym}
\usepackage{amsmath,amssymb,exscale}
\usepackage{array,multicol}
\numberwithin{equation}{section}

\def\mco{\multicolumn}

\def\ov{\overline}

\def\dalemb#1#2{{\vbox{\hrule height .#2pt
         \hbox{\vrule width.#2pt height#1pt \kern#1pt
                 \vrule width.#2pt}
         \hrule height.#2pt}}}

\let\a=\alpha    
    
    \let\p=\pi

\let\F=\Phi

 \def\bd{\begin{document}} \def\ed{\end{document}}
\def\ds{\documentstyle} \let\fr=\frac \let\bl=\bigl \let\br=\bigr
\let\Br=\Bigr \let\Bl=\Bigl
\let\bm=\bibitem
\let\na=\nabla
\let\pa=\partial
\let\ov=\overline
\def\ie{{\it i.e.\ }}
\def\tr{{\mbox{\rm tr}}}
\newcommand{\be}{\begin{equation}}
\newcommand{\ee}{\end{equation}}
\newcommand{\beba}{\begin{equation}\begin{array}{lcl}}
\newcommand{\eaee}{\end{array}\end{equation}}
\newcommand{\bea}{\begin{eqnarray}}
\newcommand{\eea}{\end{eqnarray}}
\newcommand{\ba}{\begin{array}}
\newcommand{\ea}{\end{array}}
\newcommand{\td}{\tilde}
\newcommand{\norsl}{\normalsize\sl}
\newcommand{\ns}{\normalsize}
\newcommand{\refs}[1]{(\ref{#1})}
\def\simlt{\mathrel{\lower2.5pt\vbox{\lineskip=0pt\baselineskip=0pt
            \hbox{$<$}\hbox{$\sim$}}}}
\def\simgt{\mathrel{\lower2.5pt\vbox{\lineskip=0pt\baselineskip=0pt
            \hbox{$>$}\hbox{$\sim$}}}}
\def\A{{\cal A}}
\def\a{{\mathcal a}}
\def\V{{\cal V}}
\def\F{{\cal F}}
\def\p{{\mathcal \phi}}
\def\L{{\mathcal L}}
\def\M{{\mathcal M}}
\def\bD{{\ov {\rm D}}}
\def\bO{{\ov {\rm O}}}
\def\bOp{{\ov {\rm O'}}}
\def\O{{ {\rm O}}}

\title{
\vspace*{-0.8cm}
\begin{flushright}
\normalsize{CERN-TH/2002-341\\NEIP-02-010\\ CPTH-RR-083-1102\\
\texttt{hep-ph/0211409}}\\
\end{flushright}
\vspace{1cm}
\Large\textbf{Brane to bulk supersymmetry breaking and radion force at
micron distances}
\author{\large
{\bf I.~Antoniadis~$^1$\footnote{On leave of absence from CPHT,
Ecole Polytechnique, UMR du CNRS 7644.}, K.~Benakli~$^{1,2}$\footnote{Permanent address: LPTHE, Universit\'es de Paris VI et VII, UMR du CNRS 
7589.},
A. Laugier~$^{1,3}$, T. Maillard$^{1,4}$}\\ \\
\emph{$^1$CERN Theory Division
   CH--1211, Gen{\`e}ve 23, Switzerland }\\
\emph{$^2$Institut de Physique,Universit\'e de Neuch\^atel,}\\
\emph{ CH--2000 Neuch\^atel, Switzerland}\\
\emph{$^3$Centre de Physique Th{\'e}orique, Ecole Polytechnique,}\\
\emph{91128 Palaiseau, France}\\
\emph{$^4$Institut f\"ur Theoretische Physik, ETH H\"onggerberg}\\
\emph{ CH--8093\, Z\"urich, Switzerland}}}
\date{}
\begin{document}
\maketitle
\thispagestyle{empty}
\vspace*{.5cm}

\begin{abstract}
We study mediation of supersymmetry breaking in the bulk, in models with
primordial supersymmetry breaking on D-branes at the string scale, in the
TeV region. We compute the gravitino and scalar masses up to one-loop
level, as well as the radion coupling to matter. We find that the latter
mediates a model independent force at submillimeter distances that can be
tested in micro-gravity experiments for any dimensionality of the bulk.
In the case of two large dimensions, our type I string framework provides
an example which allows to stabilize the radion potential and determine
the desired hierarchy between the string and Planck scales.
\end{abstract}
\date

\newpage
\section{Introduction}

An interesting class of type I string models with large internal
dimensions \cite{Antoniadis:1990ew, aadd} is when the closed string bulk
is supersymmetric, while supersymmetry is broken on the world volume of a
particular brane configuration \cite{bsb,abl}. This framework guarantees
the absence of quadratic divergences in the cosmological constant which
is of order
$M_s^4$ with $M_s$ the string scale, without larger contributions
proportional to $M_s^2M_P^2$ with $M_P$ the Planck mass. On the
other hand, it offers two distinct possibilities for realizing the
Standard Model. (i) It can be localized on a non supersymmetric brane
configuration, in which case the string scale must be in the TeV region
to protect the gauge hierarchy \cite{aadd}. (ii) It can be located on
some supersymmetric branes at a different position, in which case the
string scale should be at intermediate energies \cite{intermediate}, since mediation of
supersymmetry breaking in the observable world will be suppressed by the
size of the bulk.

In both cases above, one important question is to understand the
mediation of supersymmetry breaking in the bulk. In this work, we address
this question and we compute the gravitino and bulk scalar masses, up to
one loop level. Our framework is a class of type I string models, where
supersymmetry breaking is due to combinations of D-branes and
orientifolds which preserve different amount of supersymmetry. The
particle spectrum on these D-branes is then non-supersymmetric, but
supersymmetry is realized non-linearly on their worldvolume which
contains a (tree-level) massless goldstino \cite{dm,abl}. These models have in
general a localized tree-level potential, arising at the disk worldsheet,
and thus non-vanishing tadpoles of NS-NS (Neveu-Schwarz) scalar fields.
In order to avoid this problem, we can introduce a tiny source of
supersymmetry breaking in the bulk, using Scherk-Schwarz boundary
conditions along the lines of Ref.~\cite{abl}, that vanishes in the
decompactification limit. However, as it will be clear later, our results
for scalar masses do not depend on this modification, since the one-loop
mediation of supersymmetry breaking comes entirely from the branes.

We find that for more than two large bulk dimensions, the
scalar masses are always lighter, of the order of $M_s^2/M_P$, as expected
by the effective field theory, while the gravitino (and other closed
string fermions) is in general much heavier, of the order of the
compactification scale $1/R$. This is because fermions acquire tree-level
masses from the Scherk-Schwarz boundary conditions, while proper loop
corrections from the brane are extremely suppressed.
The two scales coincide in the case where
the bulk is two-dimensional (2d), since $M_s^2/M_P\sim 1/R^{n/2}$ for $n$
bulk dimensions. Also, in the special case of $n=1$, both masses become
proportional to $1/R$.

An immediate consequence of our results is that in models with the string
scale in the TeV region, the radion, the universal scalar modulus whose
vacuum expectation value (VEV) determines the size of the bulk,
mediates a new force at submillimeter distances. We thus compute its
coupling to matter and find that it is comparable to gravity and depends
only on the number of extra dimensions. Therefore, this force can be
tested in tabletop experiments that test gravity at very short distances
\cite{tests}, independently of the dimensionality of the bulk. This is in
contrast to the modification of the Newton's law which is testable only
when there are two large dimensions.

Another important question is vacuum stability. Although the general
issue goes beyond the scope of this paper, we address the problem of
stabilizing the radion \cite{Arkani-Hamed:1998kx}, assuming that the dilaton VEV is fixed and thus
the string coupling. For this purpose, we study in particular the case of
$n=2$ bulk dimensions, where there are logarithmic corrections depending
on the size of the bulk \cite{ab}.\footnote{For a recent analysis in the $n=1$ 
dimensional case, see ref.\cite{Borunda:2002ra}.} 
We then compute the effective
potential and show that the compactification scale can be fixed at
values which are hierarchically different than the string scale,
providing the desired hierarchy between $M_s$ and $M_P$.

Our paper is organized as follows. In Section 2, we derive from the
effective field theory the radion mass and its coupling to matter, using
the general form of the scalar potential that we compute in the following
sections. In Section 3, we describe our framework and we construct a
class of models that generalize those of ref.~\cite{abl} and we use for
our subsequent one-loop calculations. In Section 4, we compute the
one-loop effective potential in the case of two bulk dimensions and we
fix the radion VEV. In Section 5, we compute the string one-loop
corrections to the bulk scalar masses and to the gravitino mass. Finally,
in Section 6, we present our summary and discuss concluding remarks.

\section{Radion mass and couplings}

The kinetic terms of the radion can be obtained upon dimensional
reduction of the higher dimensional bulk Einstein-Hilbert
action:
\bea
S_{\rm bulk}^{(4+n)}=\frac{M_s^{2+n}(2 \pi)^{1-n}}{g_s^2}\,
\int\, d^{4+n}x\,\sqrt{G}\, {\cal R}^{(4+n)}\, ,
\label{action}
\eea
where we consider a bulk with $n$ large extra dimensions, while the
remaining $6-n$ are compactified at the string scale. G is the
corresponding metric and ${\cal R}^{(4+n)}$ is the Ricci scalar. Here, we
assumed that the dilaton VEV is fixed and thus the string coupling
$g_s$ is constant. Parametrizing the metric
$G={\rm diag}(g,R^2{\mathbf 1})$, with $g$ the four-dimensional (4d) metric
and $R$ a common compactification radius, one finds in the string frame:
\bea
S_{\rm kin}^{(4)}=\frac{2 \pi M_s^{2+n}}{g_s^2}\,
\int\, d^4x\,\sqrt{g}R^n\, \left\{ {\cal R}^{(4)}
-n(n-1)\left(\frac{\partial R}{ R}\right)^2 \right\}\, .
\label{actkin4}
\eea
Letting now $R_0$ the VEV of $R$ and defining the radion field $\varphi$ by
\bea
R=R_0e^{\displaystyle \kappa \varphi}\qquad ;\qquad
M_P=\frac{2^{5/2} \pi M_s^{(2+n)/2}}{ g_s}R_0^{n/2}\qquad ;\qquad \frac{1}{2 \kappa^2} = 
\frac{M_P^2}{16 \pi}\, ,
\label{radiondef}
\eea
the action (\ref{actkin4}) becomes:
\bea
S_{\rm kin}^{(4)}=\int d^4x\, e^{\displaystyle n\kappa \varphi}\,
\left[\, \frac{1}{2 \kappa^2}\, {\cal R}^{(4)}
-\frac{n\, (n-1)}{2}\, (\partial\varphi)^2\, \right]\, .
\label{actkin}
\eea

The loop corrections induce a potential localized on the brane, where
supersymmetry is broken:
\bea
S_{\rm pot}^{(4)}=\int d^4x\, \Lambda(R)=
M_s^4\, \int\, d^4x\, \left\{
\frac{1}{ g_s}V_0+V_1(R)+{\cal O}\, (g_s)\, \right\}\, ,
\label{actpot}
\eea
where $V_0$ is a constant contribution originating from the disk
worldsheet and $V_1(R)$ is the one-loop correction that we compute in
Section 4. In models without NS-NS tadpoles, that we present in the next
section, the tree-level potential $V_0$ vanishes, while $V_1(R)$ behaves
in the large radius limit as:
\bea
R\to\infty:\quad V_1(R)\sim \left\{ \begin{array}{l}
{\rm constant}\quad {\rm for}\ \ n>2\\
 \ln R\qquad\quad {\rm for}\ \ n=2\\  
R\qquad\qquad {\rm for}\ \ n=1 \end{array}\right.
\label{V1as}
\eea
The constant behavior follows from the localization of the potential,
while the logarithmic and linear in $R$ corrections are characteristic in
the cases where the bulk is two- and one-dimensional, respectively, due
to the corresponding infrared growth of the propagation of massless
fields \cite{ab}.

One can now go to the Einstein frame by rescaling the four-dimensional
(4d) metric $g_{\mu\nu}\rightarrow g_{\mu\nu}e^{-n\kappa \varphi}$, which
also diagonalizes the radion and graviton kinetic terms:
\bea
S_{\rm tot}^{(4)}=\int d^4x\,
\left[\, \frac{1}{2 \kappa^2}\, {\cal R}^{(4)}+\frac {n\, (n+2)}{ 4}\,
(\partial\varphi)^2+ e^{\displaystyle -2n\kappa \varphi}\Lambda(R)\,
\right]\, .
\label{acttot}
\eea
Assuming that the full potential has a minimum at $R= R_0$, and using
  eqs.~(\ref{actpot}) and (\ref{V1as}), it follows that in the large
volume limit the radion mass is of order $M_s^2/M_P$ for any number of
dimensions $n>1$. The case of $n=1$ is special. There is an apparent
enhancement factor $\sqrt{R_0M_s}$, but as we will see in Section 4 the
mixing among Kaluza-Klein (KK) modes becomes important and after taking it
into account, one finds a value proportional to the compactification scale
$1/R$. Thus, for $n>1$ and for a string scale $M_s\simeq 1-10$ TeV, the
radion mass is of the order of $10^{-4}-10^{-6}$ eV, which corresponds to
a wavelength of the order of 1 millimeter to 10 microns. Note that in the
case of two-dimensional (2d) bulk, there is an enhancement factor of the
radion mass by $\ln R_0M_s\simeq 30$ which decreases its wavelength by
roughly an order of magnitude.

The coupling of the radion to matter can be deduced easily from the
rescaling of the metric which changes the string to the Einstein frame,
and the normalization of its kinetic term in the action (\ref{acttot}).
Considering two masses at rest, $m_1$ and $m_2$, the amplitudes
corresponding to the exchange of a graviton and of a radion of momentum $p$ are:
\bea
{2\kappa^2m_1m_2\over p^2}\qquad {\rm and}\qquad 
{n\over n+2} {2\kappa^2m_1m_2\over p^2}\, ,
\eea
respectively.
It follows that the radion couples universally as gravity, with an attractive
force of relative strength $\alpha$:
\bea
\alpha=\frac{n}{ n+2}\, ,
\label{force}
\eea
depending only on the dimensionality of the bulk $n$ \cite{Perivolaropoulos:2002pn}. 
The values of this coupling vary from
$\alpha=1/2$ to $\alpha=3/4$ for $n=2$ to $n=6$ large extra dimensions.
Moreover, in the case of $n=2$, there may be again model dependent
logarithmic corrections of the order of $(g_s/4\pi)\ln R_0M_s\simeq{\cal
O}(1)$. Such a force can be tested in microgravity experiments and should
be contrasted with the change of Newton's law due the presence of extra
dimensions that is observable only for $n=2$ \cite{tests}.  In Fig. 1, we
plot our predictions together with the present and future experimental
limits.

\begin{figure}[htb]
\includegraphics[10mm,20mm][90mm,140mm]{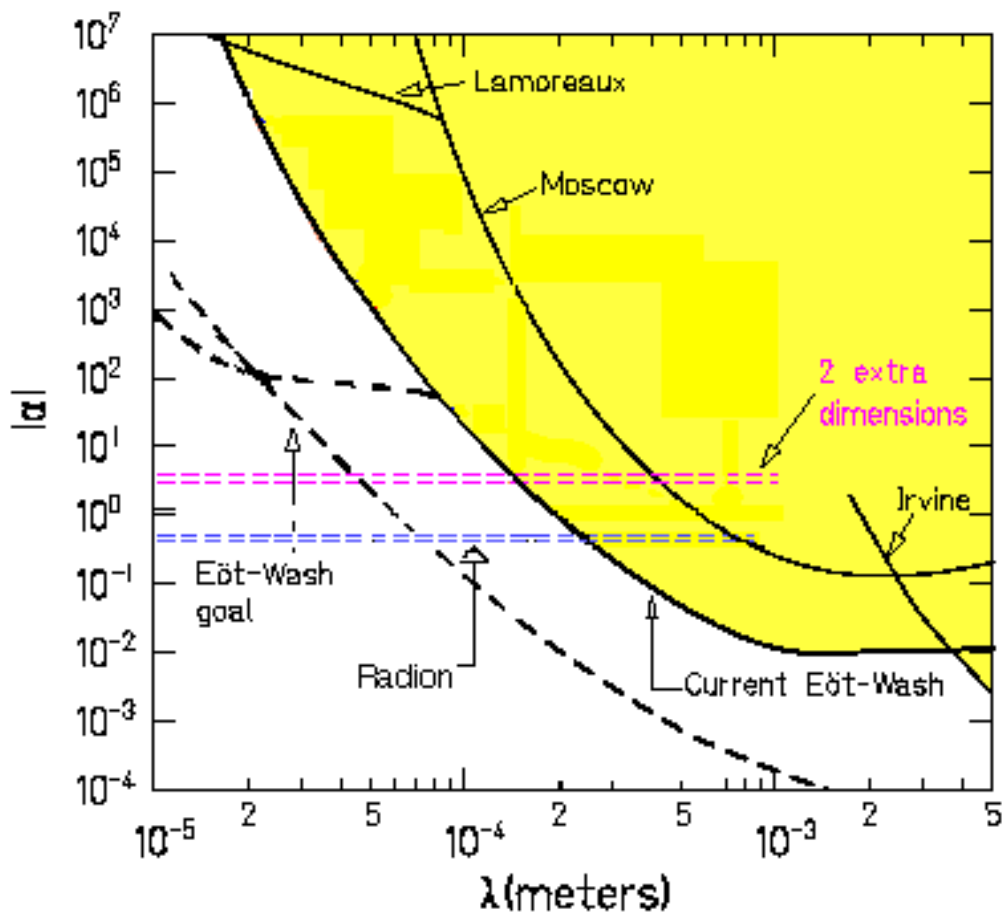}
\caption{ Present limits on non-Newtonian forces at short distances (gray
regions), compared to new forces mediated by the graviton in the case of
two large extra dimensions, and by the radion. }
\end{figure}

A final remark concerns the masses generated for other massless scalars
in the bulk. Since supersymmetry is broken only locally, on the
worldvolume of the branes, one can apply the same argument used for the
radion. All such scalars $\Phi$ are expected to receive from loop
corrections localized mass corrections, so that the relevant 4d effective
action for the zero-mode reads:
\bea
S_{\rm kin}^{(4)}=\int d^4x\, \frac{1}{ 2}
\left[\, (R_0M_s)^n \, (\partial\Phi)^2
+cM_s^2\Phi^2\, \right]\, ,
\label{scalars}
\eea
where the kinetic term comes from the bulk, as in eq.~(\ref{actkin4}),
while the mass term is localized and is fixed by the string scale, up to
a numerical constant $c$, in analogy with the radion potential
(\ref{actpot}). It follows that the physical scalar mass is suppressed by
the volume of the bulk, and is thus of the order of $M_s^2/M_P$. The
couplings of these scalars to matter from our world brane are in general
model dependent and may lead to violations of the equivalence principle
at short distances that could be measured experimentally.

\section{Model building}

To illustrate the discussion of Section 2 we will consider some 
type IIB string orientifold examples which involve compactification on a
six-dimensional space, where $n$ dimensions 
have a common large compactification radius $R>>l_s\equiv M_s^{-1}$,
while the remaining $6-n$ have a string scale size and do not play a major 
role. 

The simplest cases on which the desired computations can be 
carried out consist of toroidal compactifications with the adjunction of 
a certain number of orientifold planes and D-branes. These come in
different  kinds (see Table 1) as they are distinguished by the charges
they carry, both Neveu-Schwarz--Neveu-Schwarz (NS-NS), or equivalently
their tension, and Ramond-Ramond (RR) type. Here, we will consider for
simplicity D-branes and orientifold planes of the same dimensionality
$p$. The generalization to the case of several types extended in
different internal dimensions is straightforward.

\begin{table}[t]
\begin{center}
\renewcommand{\arraystretch}{0.3}
\begin{tabular}{  | c || c | c | c | c || c | l |}
\hline &  \mco{4}{c||}{}  &\mco{2}{c|}{}
\\ & \mco{4}{c||}{Orientifolds}&\mco{2}{c|}{D-branes}\\
   &  \mco{4}{c||}{}  &\mco{2}{c|}{}
\\ \hline\hline & & & & & & \\
   Symbol  & \, \, ${\O}^-_p$ \, \, &  \, \,${\bO}^-_p$  \, \, &  \, 
\,${\O}^{+}_p$ \, \,
   & \, \, ${\bO}^{+}_p$ \, \, & \, \, D$p$ \, \, & \, \,  ${\bD}p$ 
\, \\
   & & & & & & \\
\hline  & & & & & & \\ RR charge  & $-$ &  + & + & $-$ & +& \, \, $-$ \\
   & & & & & & \\ \hline  & & & & & & \\
   NS--NS charge & $-$& $-$ &+ & + &+ & \,  \, + \\  & & & & & & \\
   \hline & & & & & & \\
   Number & $N_p^-$& ${\bar N}_p^-$ &$N_p^+$ & ${\bar N}_p^+$ & $n_p$ & \,  \, 
 ${\bar n}_p$ \\  & & & & & & \\
   \hline  & & & & & & \\
   Supersymmetries & $Q$& ${\bar Q}$ &$Q$ & ${\bar Q}$ & $Q$,\,  
${\bar Q}_{\rm NL}$ & ${\bar Q}$,\, $Q_{\rm NL}$ \\  & & & & & & \\
   \hline
\end{tabular}
\end{center}
\caption{The RR and NS--NS charges of orientifolds and D-branes.
}\end{table}

In our conventions, D-branes (anti-D-branes) have positive (negative) RR
charge and positive tension, while there are two types of orientifolds:
$\O^-$-planes with negative tension and $\O^+$-planes with positive
tension. Thus, $\O^+$-planes (${\bO}^+$-planes) have 
the same
quantum numbers as D-branes ($\bD$-branes), while $\O^-$-planes
($\bO^-$-planes) have opposite ones. In Table 1, we also display the
supersymmetries preserved. D-branes and both types of orientifolds
preserve the same amount of supersymmetry $Q$, which amounts half of the
bulk supersymmetry, while their conjugates $\bD$ and ${\bO},
{\bO}^\pm$ preserve the other half ${\bar Q}$. On the other hand,
D-branes ($\bD$-branes) realize also the other half ${\bar Q}$ ($Q$) in a
non-linear way, indicated with a subscript NL in the table.

Let us specialized for definiteness to the case of two transverse
dimensions with D7-branes and orientifold 7-planes, $p=7$. The total RR
charge of the system:
\bea
Q^{RR}= -8N_p^- + 8{\bar N}_p^-+ 8N_p^+ -8{\bar N}_p^+ +n_p -{\bar n}_p
\eea
is constrained to vanish, $Q^{RR}=0$, due to the flux conservation 
inside the compact space. On the other hand, the total NS-NS charge of
the  system is given by:
\bea
Q^{NS}= -8N_p^- - 8{\bar N}_p^-+ 8N_p^+ +8{\bar N}_p^+ +n_p +{\bar n}_p
\eea
is not required to vanish in general by consistency. However, a
non-vanishing $Q^{NS}$ implies a non-vanishing tadpole for the
corresponding NS-NS scalar (in our case the dilaton), accompanied by a
non-trivial (usually runaway) tree-level potential. Following the method
of ref.~\cite{abl}, we will start by constructing models with no tadpoles,
that correspond to consistent non-supersymmetric string vacua in flat
space at the tree-level. Since we are interested to stabilize the
compactification radius at values much larger than the string length, we
will specialize to the case of a two-dimensional bulk and compute the
one-loop radion potential which behaves logarithmically in the large
radius limit, according to eq.~(\ref{V1as}). However, we will be unable
to find consistent solutions in this limit. 

On the other hand, since the vacuum is non-supersymmetric, higher
loop corrections are expected to generate in any case a non-vanishing
dilaton potential, and thus dilaton tadpole, related to the familiar
problem of the cosmological constant, for which we have nothing new to
say. We will therefore allow, in a second step, the presence of a
potential, and thus of an uncanceled dilaton tadpole, already at the
tree-level. We will then show that one can stabilize the radion
and determine the desired hierarchy by combining the tree-level
contribution with the logarithmic one-loop corrections, in a way
analogous to the no-scale mechanism or the so-called inverse hierarchy of
softly broken supersymmetry \cite{Coleman:jx}.

As we already stated in the introduction, we will assume that the dilaton
VEV is fixed and ignore the presence of a cosmological constant at the
minimum of the potential, focusing only on fixing the radion. Finally,
the compactification will proceed in two steps: first down to 
$p+1=10-n$ and then down to four. Here, we will describe in detail only
the  first step, while the second is trivial for toroidal or orbifold
compactifications.

\subsection{Susy breaking along one extra dimension}

In \cite{abl}, a construction of consistent type I vacua with
spontaneously broken supersymmetry was presented. The main ingredients
of the construction are the following:

\begin{itemize}

\item The ten-dimensional type IIB  string is compactified on $X \otimes
{\M_4}$ with an orientifold projection on the worldsheet. Here ${\M_4}$
is  the four-dimensional Minkowski space and $X$ is a compact manifold
which contains a segment $S^1/Z_2$ used to break  supersymmetry in the
bulk. The breaking is achieved by a choice of different periodicity
conditions around $S^1$ for the fermions and bosons in the same
supermultiplets (Scherck-Schwarz mechanism). In our construction we
choose all the bosons periodic, while the fermions are anti-periodic.

\item At each of the end-points of the segment $S^1/Z_2$, the orientifold
projection introduces  an orientifold plane. The Scherck--Schwarz (SS)
boundary conditions require the two orientifolds to preserve different
halves of the bulk supersymmetry, which is achieved for instance by having
the orientifolds to be of the same type and to carry opposite
RR charges.

\item In order to insure the absence of NS-NS tadpoles,
we require that the orientifolds carry negative NS-NS charges which are
cancelled by adding D-branes. The cancellation of the total RR charge on
the other hand implies that the D-branes appear in pairs of
brane--anti-brane.

\item Two possible consistent models are found. In the first case
(named ``brane supersymmetry") the branes are placed on top of
orientifolds preserving the same supersymmetries. The massless modes
localized at the boundaries are then degenerate between bosons and
fermions and form supermultiplets. A second case (named ``brane
supersymmetry breaking") is obtained by putting the branes on top of
orientifolds preserving different supersymmetries. The massless modes
left on the branes are not anymore degenerate between bosons and
fermions but the effective field theory has a non-linearly realized
supersymmetry. These properties are easily understandable by inspection
of Table 1.

\end{itemize}

\subsection{Models with two compact dimensions}

In this section, we present an extension of the above construction
with branes and anti-branes located at different points of
a two-dimensional compactification plane instead of a line. We will
consider two cases, where the non-periodic boundary conditions for bulk
fermions are imposed along one or both  directions. We will denote the
two classes of models as $(I)$ and $(II)$.

The one-loop partition function contains four terms accounting for the
contributions of the torus ($\cal T$), the Klein bottle ($\cal K$), the
annulus ($\cal A$) and the M\"obius  strip ($\cal M$). For simplicity, we
will consider an orthogonal torus with equal radii $R$. Also, below we
present only the simplest models defined effectively in eight dimensions.
The generalization to lower dimensions, with less supersymmetry in the
bulk, and various types of D-branes is straightforward and can be easily
done along the lines of refs.~\cite{abl}.

The torus contribution is given by (see \cite{abl} for the integration
measures):
\bea
{\cal T}_I&=& \left[
E'_0(V_8{\bar V}_8+S_8{\bar S}_8)+O'_0(I_8{\bar I}_8+C_8{\bar C}_8)
\right. \nonumber \\ \nonumber &&- \left.
E'_{\frac{1}{2}}(V_8{\bar S}_8+S_8{\bar V}_8)-O'_{\frac{1}{2}}(I_8{\bar
C}_8+C_8{\bar I}_8) \right] \, \sum_{m,n} Z_{m,n}
\eea
in the case where the SS boundary condition is used along one direction,
and by:
\bea
{\cal T}_{II} &=& (E'_0E'_0 + O'_{0}O'_{0})(V_8{\bar V}_8+S_8{\bar S}_8)
\nonumber\\ &&+
(E'_0 O'_{0}+ O'_{0}E'_0) (I_8{\bar I}_8+C_8{\bar C}_8) \nonumber\\ &&-
(E'_{1/2}E'_{1/2}+O'_{1/2}O'_{1/2})(V_8{\bar S}_8+S_8{\bar V}_8)\nonumber\\ &&-
(E'_{1/2}O'_{1/2}+O'_{1/2}E'_{1/2})(I_8{\bar C}_8+C_8{\bar I}_8)
\eea
for the case where the SS deformation is used along both directions. Our
notations are the same as in ref.~\cite{abl}. $I_8$, $V_8$, $S_8$ and
$C_8$ are the $SO(8)$ characters corresponding to the conjugacy classes
of the identity, vector, spinor and conjugate spinor, respectively. Their
expressions in terms of theta-functions are given by:
\bea
I_8={\theta_3^4+\theta_4^4\over 2\eta^4}\quad
V_8={\theta_3^4-\theta_4^4\over 2\eta^4}\quad
S_8={\theta_2^4+\theta_1^4\over 2\eta^4}\quad
C_8={\theta_2^4-\theta_1^4\over 2\eta^4}\, ,
\eea
where $\eta$ is the Dedekind eta-function. $Z_{m,n}$ is the
one-dimensional torus partition function,
while $E'_a$ ($O'_a$) denotes the partition function restricted to even
(odd) windings $n$ and shifted  momenta  $m+a$:
\bea
Z_{m,n}={1\over |\eta|^2}q^{{\alpha^\prime \over 4}
({m\over R}+{nR\over\alpha^\prime})^2}
{\bar q}^{{\alpha^\prime \over 4} ({m\over R}-{nR\over\alpha^\prime})^2}\quad
E'_a=\sum_{m,n}Z_{m+a,2n}\quad O'_a=\sum_{m,n}Z_{m+a,2n+1}\, ,
\eea
with $\alpha^\prime=l_s^2$, $q=e^{2 i\pi\tau}$ and $\tau$ the complex
modulus of the worldsheet torus.

The Klein bottle contribution can be written as:
\bea
{\cal K}_I&=& \frac{1}{2}\left( (V_8-S_8)\sum_{m_1} \tilde{Z}_{2m_1}
+(I_8-C_8)
\sum_{m_1} \tilde{Z}_{2m_1+1}\right)\, \sum_{m_2} \tilde{Z}_{m_2}  
\nonumber\\
{\cal K}_{II}&=&\frac{1}{2}(V_8-S_8)\!\!\!\!\sum_{m_1+m_2\,even}
\tilde{Z}_{m_1}\tilde{Z}_{m_2}
+\frac{1}{2}(I_8-C_8)\!\!\!\!\sum_{m_1+m_2\,odd}\tilde{Z}_{m_1}
\tilde{Z}_{m_2}
\nonumber
\eea
in the direct (open string) channel where
\bea
{\tilde Z}_n={q^{\frac {n^2 R^2}{2 \alpha^\prime}}\over\eta}\, ,
\eea
 and
\bea
\tilde{\cal K}_{I}&=& \frac{2^5}{2}\alpha^\prime R^{-2}
(V_8\sum_{n_1} {Z}_{2n_1}-S_8\sum_{n_1} {Z}_{2n_1+1})
\sum_{n_2} {Z}_{2n_2} \nonumber\\
\tilde{\cal K}_{II}&=&\frac{2^5}{2}\,\alpha^\prime R^{-2} \,
\left(V_8 { Z}_{ee}-S_8 { Z}_{oo}\right)
\nonumber
\eea
in the transverse (closed string) channel. Here again we used the
notation:
\bea
Z_m={q^{{\alpha^\prime \over 4}({m\over R})^2}\over\eta}
\eea
while ${Z}_{ee}$ and ${Z}_{oo}$ are the two-dimensional 
partition function sums, restricted to (even, even) and (odd, odd)
momenta, respectively.

The remaining two contributions describe the content of the open
string sector in these models. The annulus vacuum amplitude, in the
transverse channel, can be expressed as:
\bea
\tilde{\cal A} &=& 
\frac{2^{-5}}{2}\frac {\alpha^\prime}{R^2}
(\alpha_1^2V_8 {Z}_{ee}+\alpha_2^2V_8 
{Z}_{oo}-\beta_1^2S_8 {Z}_{ee}-\beta_2^2S_8 {Z}_{oo}+\gamma_1^2V_8
{Z}_{eo}\nonumber 
\\ && +\gamma_2^2V_8 {Z}_{oe}- 
\delta_1^2S_8 {Z}_{eo}-\delta_2^2S_8 {Z}_{oe})\, ,
\eea
where $\alpha_j,\,\beta_j,\,\gamma_j,\,\delta_j\,\,(j=1,2)$ are
the boundary reflection coefficients. As usual, an appropriate
parametrization in terms of the Chan-Paton factors is given by:
\bea
\label{Parametric}
&&\alpha_1=n_{00}+n_{\pi \pi}+{\bar n}_{\pi \pi}+n_{\pi 0}+{\bar n}_{\pi 0}
+{\bar n}_{0 0}+n_{0 \pi}+{\bar n}_{0 \pi} \nonumber \\
&&\alpha_2=n_{00}+n_{\pi \pi}+{\bar n}_{\pi \pi}-n_{\pi 0}-{\bar n}_{\pi 0}
+{\bar n}_{0 0}-n_{0 \pi}-{\bar n}_{0 \pi} \nonumber \\
&&\beta_1=n_{00}+n_{\pi \pi}-{\bar n}_{\pi \pi}+n_{\pi 0}-{\bar n}_{\pi 0}
-{\bar n}_{0 0}+n_{0 \pi}-{\bar n}_{0 \pi} \nonumber \\
&&\beta_2=n_{00}+n_{\pi \pi}-{\bar n}_{\pi \pi}-n_{\pi 0}+{\bar n}_{\pi 0}
-{\bar n}_{0 0}-n_{0 \pi}+{\bar n}_{0 \pi} \nonumber \\
&&\gamma_1=n_{00}-n_{\pi \pi}-{\bar n}_{\pi \pi}+n_{\pi 0}+{\bar n}_{\pi 0}
+{\bar n}_{0 0}-n_{0 \pi}-{\bar n}_{0 \pi} \nonumber \\
&&\gamma_2=n_{00}-n_{\pi \pi}-{\bar n}_{\pi \pi}-n_{\pi 0}-{\bar n}_{\pi 0}
+{\bar n}_{0 0}+n_{0 \pi}+{\bar n}_{0 \pi} \nonumber \\
&&\delta_1=n_{00}-n_{\pi \pi}+{\bar n}_{\pi \pi}+n_{\pi 0}-{\bar n}_{\pi 0}
-{\bar n}_{0 0}-n_{0 \pi}+{\bar n}_{0 \pi} \nonumber \\
&&\delta_2=n_{00}-n_{\pi \pi}+{\bar n}_{\pi \pi}-n_{\pi 0}+{\bar n}_{\pi 0}
-{\bar n}_{0 0}+n_{0 \pi}-{\bar n}_{0 \pi}
\eea
where $n_{y_1\,y_2}$ (${\bar n}_{y_1\,y_2}$) represents the number of
branes (anti-branes) sitting at the point $(y_1\, y_2)$ with
$y_{1,2}=0$ or $\pi R$.

With this parametrization, the direct channel 
annulus amplitude takes the form:
\bea
{\cal A} &=& (V_8-S_8)\left\{
\frac{n_{00}^2+{\bar n}_{00}^2+n_{\pi 0}^2+{\bar n}_{\pi 0}^2+n_{0
\pi}^2+{\bar n}_{0 \pi}^2+n_{\pi \pi}^2+{\bar n}_{\pi
\pi}^2}{2}\sum_{m_1,m_2}{\tilde Z}_{m_1}{\tilde Z}_{m_2} \right.
\nonumber \\
&+& (n_{00}n_{\pi \pi}+n_{0 \pi}n_{\pi 0}+{\bar n}_{00}{\bar n}_{\pi
\pi}+{\bar n}_{0 \pi}{\bar n}_{\pi 0})\sum_{m_1,m_2}{\tilde Z}_{m_1+1/2}
{\tilde Z}_{m_2+1/2} \nonumber \\     &+& (n_{00}n_{\pi 0}+{\bar
n}_{00}{\bar n}_{\pi 0}+n_{0 \pi}n_{\pi \pi}+{\bar n}_{0 \pi}{\bar
n}_{\pi \pi})\sum_{m_1,m_2}{\tilde Z}_{m_1+1/2} {\tilde Z}_{m_2}
\nonumber \\ &+& \left.
(n_{00}n_{0 \pi}+n_{\pi 0}n_{\pi \pi}+{\bar n}_{00}{\bar
n}_{0 \pi}+{\bar n}_{\pi 0}{\bar n}_{\pi \pi})\sum_{m_1,m_2} {\tilde
Z}_{m_1}{\tilde Z}_{m_2+1/2}\right\}
 \nonumber \\ &+&(I_8-C_8)\left\{ ({\bar
n}_{00}n_{00}+{\bar n}_{\pi 0}n_{\pi 0}+{\bar n}_{0 \pi}n_{0 \pi}+{\bar
n}_{\pi \pi}n_{\pi \pi})\sum_{m_1,m_2} {\tilde Z}_{m_1}{\tilde Z}_{m_2}
\right. \nonumber \\&+& 
({\bar n}_{00}n_{\pi \pi}+{\bar n}_{0 \pi}n_{\pi 0}+n_{0
\pi}{\bar n}_{\pi 0}+n_{00}{\bar n}_{\pi \pi})\sum_{m_1,m_2}{\tilde
Z}_{m_1+1/2}{\tilde Z}_{m_2+1/2} \nonumber \\ &+& (n_{00}{\bar n}_{0
\pi}+{\bar n}_{00}n_{0 \pi}+n_{\pi 0}{\bar n}_{\pi \pi}+{\bar n}_{\pi
0}n_{\pi \pi})\sum_{m_1,m_2} {\tilde Z}_{m_1}{\tilde Z}_{m_2+1/2}
\nonumber \\ &+& \left.
(n_{00}{\bar n}_{\pi 0}+{\bar n}_{00}n_{\pi 0}+{\bar
n}_{0 \pi}n_{\pi \pi}+n_{0 \pi}{\bar n}_{\pi \pi})\sum_{m_1,m_2} {\tilde
Z}_{m_1+1/2}{\tilde Z}_{m_2})\right\}\, .
\nonumber
\eea

The M\"obius strip partition function in the transverse channel is
then easily found to be:
\bea
\tilde{\cal M}_I&=& 
-\frac {\alpha^\prime}{R^2}\sum_{n_1,n_2}(\alpha_1\varepsilon
\hat{V}_8 {Z}_{2n_1} {Z}_{2n_2} 
- \delta_2\hat{S}_8 {Z}_{2n_1+1} {Z}_{2n_2})
\nonumber\\
\tilde{\cal M}_{II}&=&-\frac {\alpha^\prime}{R^2}(\alpha_1\varepsilon
\hat{V}_8 {Z}_{ee}-\beta_2\hat{S}_8 {Z}_{oo})\, ,
\nonumber
\eea
which becomes in the direct channel:
\bea
{\cal M}_I &=& - \frac{1}{2} \sum_{m_1,m_2}(\alpha_1\varepsilon
\hat{V}_8 - (-1)^{m_1}
\delta_2\hat{S}_8) {\tilde Z}_{m_1} {\tilde Z}_{m_2}
\nonumber\\
{\cal M}_{II} &=& -\frac{1}{2}\sum_{m_1,m_2}(\,\,\alpha_1\varepsilon
\hat{V}_8
- (-1)^{m_1+m_2}\beta_2\hat{S}_8) {\tilde Z}_{m_1} {\tilde Z}_{m_2}\, .
\nonumber
\eea
The parameter $\varepsilon$ is a sign ambiguity which we discuss below.
It is introduced only in the coefficient of $\hat{V}_8$ because of the
positivity of $\alpha_1$ following the parametrization
(\ref{Parametric}). As we explain below, it reflects the freedom of
introducing one of the two types of orientifolds, $\O^-$ or $\O^+$,
defined in Table 1.

The integral over the modular parameter 
$\tau = \tau_1 + i\tau_2$ in the torus amplitude ${\cal T}$ is performed
over the fundamental domain 
\bea 
{\cal F}:\quad -\frac {1}{2} \leq \tau_1 \leq \frac {1}{2}\, ,\quad 
\tau_2 \geq 0\, ,\quad |\tau| \geq 1 \, .
\nonumber
\eea
The direct (open string) channel amplitudes $ {\cal K}$, $ {\cal A}$ and
$ {\cal M}$ are integrated over $t \in [0,\infty)$. The  corresponding
expressions in the transverse  channel $ {\tilde {\cal K}}$, 
${\tilde {\cal A}}$ and ${\tilde {\cal M}}$ are integrated over
 $l \in [0,\infty)$, obtained by the following transformations:
\bea 
\tilde{\cal K} &:& \qquad\qquad \qquad 2 it  \ \, \,
 {\longrightarrow}  \qquad
 \frac {i} { 2 t}\equiv il \\
\tilde{\cal A} &:& \qquad\qquad \qquad \frac {it}{2} \ \ \,
{\longrightarrow}  \qquad
\frac{2i}{t} \equiv il \\
\tilde{\cal M} &:& \qquad\qquad {it \over 2}+{1 \over 2} \ \ \,
{\longrightarrow}   \qquad
{i \over 2t}+{1 \over 2} \equiv il + {1 \over 2} \ . 
\label{dirtotrans}
\eea
The $t$-dependent parameters in the l.h.s. of the arrows appear as
argument of the $\theta$ and $\eta$ functions
in the corresponding vacuum amplitudes ${\cal K}$, ${\cal A}$ and 
${\cal M}$, while the $l$-dependent ones on the r.h.s.
are the corresponding arguments in $ {\tilde {\cal K}}$, 
${\tilde {\cal A}}$ and ${\tilde {\cal M}}$. Finally, the hat on the
characters of ${\cal M}$ and ${\tilde {\cal M}}$ stands as usual for
the translation by 1/2 according to eq.~(\ref{dirtotrans}).

The exact particle content of the models is fixed after imposing three
consistency conditions: i) the absence of tachyons, ii) the absence of RR
tadpoles and iii)  the absence of tree-level NS-NS tadpoles. Actually,
the last requirement can be satisfied anly for $\varepsilon=+$. However,
since in the next section we will relax this condition, we will discuss
here both cases with the understanding that NS-NS tadpoles do not vanish
for $\varepsilon=-$. The first requirement is satisfied if there are no
pairs of brane--anti-brane at short distance (shorter than the string
length). This is achieved by taking $n_{y_1\,y_2}{\bar n}_{y_1\,y_2}=0$,
for all $y_1,\,y_2$, and by considering the compactification radii
transverse to the branes large enough.

The resulting models\footnote{Here, we restrict our analysis to branes
and anti-branes  on top of orientifolds. Introduction of Wilson lines
would allow to have  new models with branes in the bulk.} are:

\begin{itemize}
\item Model A, where all branes (anti-branes) are on top of orientifolds
(anti-orientifolds). In this case, supersymmetry is not broken locally
and the massless open string states present a fermion-boson degeneracy
at tree-level and form supersymmetric multiplets.  This model can be
obtained by:
\bea
{\rm Model\,\,\, A}\, I:&& {\bar n}_{00}=n_{\pi 0}={\bar n}_{0 \pi}=
n_{\pi \pi}=0 \nonumber \\ && n_{00}+n_{0 \pi}={\bar n}_{\pi 0}+{\bar
n}_{\pi \pi}=16
  \\
{\rm Model\,\,\, A}\, II:&& {\bar n}_{00}=n_{\pi 0}= n_{0 \pi}=
{\bar n}_{\pi \pi}=0 \nonumber \\ && n_{00}+n_{\pi \pi}={\bar n}_{\pi 
0}+{\bar n}_{0 \pi}=16
\eea

\item Model B, where all branes (anti-branes) are on top of
anti-orientifolds  (orientifolds). In this case, supersymmetry is broken
locally at the positions of the branes and anti-branes, and the massless
open string spectra do not have a fermion-boson degeneracy
but satisfy a non-linear supersymmetry \cite{abl}. This model
corresponds to:
\bea
{\rm Model\,\,\, B}\, I:&& n_{00}=n_{0 \pi}={\bar n}_{\pi 0}={\bar 
n}_{\pi \pi}=0 \nonumber \\ && n_{\pi 0}+n_{\pi \pi}={\bar 
n}_{00}+{\bar n}_{0 \pi}=16
  \\
{\rm Model\,\,\, B}\, II:&& n_{00}=n_{\pi \pi}={\bar n}_{\pi 0}={\bar 
n}_{0 \pi}=0\nonumber\\ && 
{\bar n}_{00}+{\bar n}_{\pi \pi}=n_{\pi 0}+ n_{0 \pi}=16
\eea

\item Model C with mixed configuration, where some of the branes are on
top of orientifolds and have fermion-boson degenerate massless spectra,
while others sit on top of anti-orientifolds and have a non-linearly
realized supersymmetry on their worldvolume and non supersymmetric
spectrum. We can split these models to two classes: (a) {\bf C $Ia$} and
{\bf C $IIa$}, having two 
non-supersymmetric and two supersymmetric sectors; these can be
either on neighboring or on opposite edges of the square of the two
compact dimensions (see Fig. 2);
(b) {\bf C $Ib$} and {\bf C $IIb$}, having three or one 
supersymmetric sectors.
Examples of such models are obtained as:
\bea
{\rm Model\,\,\, C} \,\,Ia\, ,\,IIa :&& n_{00}=n_{\pi 0}={\bar n}_{0 
\pi}={\bar n}_{\pi \pi}=0 \nonumber \\ && n_{0 \pi}+n_{\pi \pi}={\bar 
n}_{00}+{\bar n}_{\pi 0}=16 \\
{\rm Model\,\,\, C} \,\,Ib\, ,\,IIb :&& {\bar n}_{00}={ n}_{\pi 
0}=n_{0 \pi}=n_{\pi \pi}=0 \nonumber \\ && n_{00}={\bar n}_{\pi 0}+{\bar 
n}_{0 \pi}+{\bar n}_{\pi \pi}=16
\eea
These models can provide a natural setting
for further studies of the mediation of supersymmetry breaking in brane
models.
\end{itemize}

\begin{figure}[htb]
\centering
\epsfxsize=4.0in
\epsfbox{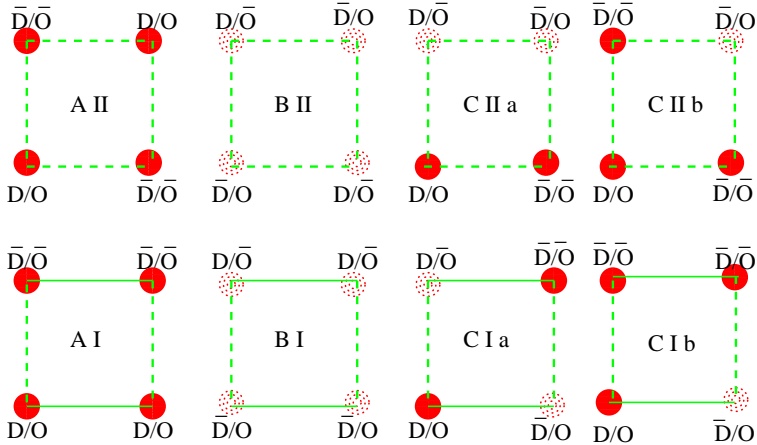}
\caption{ The supersymmetric, non-supersymmetric and mixed models. The
dotted lines denote directions along which non-periodic  boundary
conditions are imposed on the bulk fermions. The plain (dotted) disks
stand for locations of supersymmetric (non-supersymmetric) branes. }
\end{figure}

A simple inspection of the annulus and M\"obius amplitudes in the direct
channel shows that, depending the sign ambiguity $\varepsilon=\pm$, the
Chan Paton charges of the gauge fields are antisymmetrized or
symmetrized, respectively. As a result, in the case $\varepsilon =+$,
the gauge group is $\prod_i SO(n_i)$ with $n_i$
the non-vanishing Chan-Paton factors (here $n_i$ denote collectively also
${\bar n}_i$). In supersymmetric massless sectors, the bosons and
fermions belong to vector supermultiplets. In non-supersymmetric
massless sectors, the bosons remain in the adjoint (antisymmetric) of
$SO(n)$ representation, while fermions are in general in the symmetric
ones. In contrast, in the $\varepsilon =-$ case, the gauge group is 
$\prod_i Sp(n_i)$ and in non-supersymmetric
massless sectors, the bosons are in the adjoint (symmetric) while
fermions are in antisymmetric representations.

In the decompactification limit $R \rightarrow \infty$, 
supersymmetry is restored in the closed string sector while it is broken 
on the worldvolume of the non-supersymmetric branes. This is reflected 
in the non-vanishing of the M\"obius amplitude which accounts for the
contribution of superstrings stretched between branes and orientifolds
and provides the only source for supersymmetry breaking that remains in
this limit. As we mentioned before, these models can be used as building
blocks for more general constructions in lower dimensions with less
supersymmetry, several types of branes and chiral matter
representations.

\section{One-loop effective potential and radius stabilization}
\subsection{Explicit examples}

In this section, we compute the one-loop cosmological constant $\Lambda$
in the limit of large transverse radii $R>>l_s$. Since the dilaton is
considered constant, we will be interested only in the $R$
dependence of the effective potential. The expressions here are given in
the string frame.

We first consider the contribution from the torus amplitude (closed
string sector) $\Lambda_c$. In the  limit $R \rightarrow \infty$, the
winding modes decouple and we are left over with the zero winding sector:
\bea
\Lambda_{c}\!\!\!\! &\stackrel{R \rightarrow \infty}{\sim}&
\!\!\!\! \int_{\mathcal{F}}\frac{d^{2}\tau}{\tau_{2}^{5}}\sum_{m_1,m_2}
\left[{Z}_{m_{1},0} {Z}_{m_{2},0}(|V_{8}|^{2} + |S_{8}|^{2}) -
{Z}_{m_{1}+\frac{1}{2},0} {Z}_{m_{2}+\frac{\epsilon}{2},0}
(V_{8}\bar{S}_{8} + S_{8}\bar{V}_{8})\right]
     \label{eq:tore1} \nonumber\\
      &  \stackrel{R \rightarrow \infty}{\sim}&
      \frac {2}{R^{8}}\!\!\int_{\mathcal{F}}\!\!
\frac{d^{2}\tau}{\tau_{2}^{5}}\sum_{m_1,m_2}
      \left|\frac{\theta_{2}^{4}}
      {2\eta^{12}}\right|^{2}(\!\tau_{1}\! +\! i\tau_{2}\! R)
\left(e^{-\pi(m_{1}^{2}+m_{2}^{2})\tau_{2}}
-e^{-\pi [(m_{1}+\frac{1}{2})^{2}+(n_{2}+
\frac{\epsilon}{2})^{2}]\tau_{2}}\right)
  \nonumber \\
      & \stackrel{R \rightarrow \infty}{\sim}&
      \frac {2^{7}}{R^{8}}\int_{0}^{\infty}
      d\tau_{2}\tau_{2}^{4}\sum_{n_1,n_2}
      (1 - (-1)^{n_{1}+\epsilon n_{2}})e^{-\pi \tau_{2}(n_{1}^{2}
+n_{2}^{2})}  \nonumber\\
      & \stackrel{R \rightarrow \infty}{\sim} & (n_{b}^{c} - n_{f}^{c})
\frac {1}{R^{8}}
\end{eqnarray}
where $\epsilon=0,1$ for the cases $I$ and $II$, respectively, while in
the third line we performed a change of variables $\tau_2\to 1/\tau_2$
and a Poisson resummation in $m_1$ and $m_2$ to $n_1$ and $n_2$. Here,
$n_{b}^{c}$ and $n_{f}^{c}$ are the number of massless bosons and
fermions from the closed string sector. Note that the power of $1/R$
depends actually on the number of non-compact spacetime dimensions, so
that in four-dimensional compactifications the $1/R^8$ becomes $1/R^4$.

We turn now to the contribution $\Lambda_o$ of the open (and closed
unoriented) string sector, coming from the remaining three surfaces 
${\cal K+ A+ M}$. Using that the absence of tachyon needs
$\alpha_1^2+\alpha_2^2-\beta_1^2-\beta_2^2+\gamma_1^2+\gamma_2^2- 
\delta_1^2-\delta_2^2 =0 $ 
and that the cancellation of the RR total charge implies $\beta_1=0$,
we find:
\bea
\label{GCKAM}
\Lambda_o|_{I} &=& -\frac{1}{(4\pi^2\alpha')^{d/2}}
\int_0^{\infty}\frac{dl}{l} \nonumber \\
&&\!\!\!\!\!\! \left[ \frac{\theta_2^4}{\eta^{12}}(il)
(\frac{\alpha_1^2+ 2^{10}}{2^8}\theta_3^2
+\frac{\alpha_2^2-\beta_2^2}{2^8}\theta_4^2-
\frac{\alpha_1^2+\alpha_2^2-\beta_2^2+2^{10}}{2^8}\theta_3\theta_4)
(\frac{iR^2}{2\alpha' l})\right.
\nonumber \\
&&-\left. \frac{\hat{\theta}_2^4}{\hat{\eta}^{12}}(il+\frac{1}{2})
(\frac{2^6}{2^8}\alpha_1\varepsilon\theta_3^2-
\frac{2^6}{2^8}\delta_2\theta_3\theta_4)
(\frac{iR^2}{2\alpha'l})\right]\, ,  \\
\Lambda_o|_{II} &=&-\frac{1}{(4\pi^2\alpha')^{d/2}}
\int_0^{\infty}\frac{dl}{l}\nonumber \\ 
&&\!\!\!\!\!\! \left[ \frac{\theta_2^4}{\eta^{12}}(il)
(\frac{\alpha_1^2+ 2^{10}}{2^8}\theta_3^2+
\frac{\alpha_2^2-\beta_2^2-2^{10}}{2^8}\theta_4^2-
\frac{\alpha_1^2+\alpha_2^2-\beta_2^2}{2^8}\theta_3\theta_4)
(\frac{iR^2}{2\alpha'l})\right.\nonumber 
\\ &&-\left. 
\frac{\hat{\theta}_2^4}{\hat{\eta}^{12}}(il+\frac{1}{2})(
\frac{2^6}{2^8}\alpha_1\varepsilon\theta_3^2-
\frac{2^6}{2^8}\beta_2\theta_4^2)
(\frac{iR^2}{2\alpha'l})\right]\, ,
\label{GCKAMII}
\eea
where $d$ is the number of non-compact dimensions.
In the limit $R \rightarrow \infty$, the resulting potential presents a
logarithmic divergence. This arises from the integration region $l
\rightarrow \infty$ which corresponds to the ultraviolet (UV) limit in the
open string channel. In fact, for finite $R$, in the limit $l\to\infty$,
the integrand of the expressions (\ref{GCKAM}) and (\ref{GCKAMII}) is
exponentially suppressed as a result of a cancellation between the two
$\theta_3^2$ terms, when $\alpha_1=32$ and $\varepsilon=+$ which is the
condition for vanishing of the dilaton tadpole. 

More precisely, using the properties of $\theta$-functions, one can
invert their argument $iR^2/2\alpha'l\to i2\alpha'l/R^2$ by replacing
$\theta_3\to (2\alpha'l)^{1/2}\theta_3/R$ and $\theta_{2,4}\to
(2\alpha'l)^{1/2}\theta_{4,2}/R$; one can then take the limit 
$l\to\infty$ using
their asymptotic expansions, implying $\theta_{3,4}(il)\to 1$ and
$\theta_{2}\to 0$ with
$\theta_{2}/\eta^3(il)\to 2$. As a result, one is left over with a
quadratic divergence $\sim dl$ proportional to the NS-NS dilaton
tadpole coefficient 
\bea
\frac {2^3}{2^6} \frac{1}{(4\pi^2\alpha')^{d/2}}\int^\infty dl\ :\qquad
(\alpha_1^2+2^{10}-2^6\alpha_1\varepsilon)\frac{\alpha'}{R^2}=
\frac{(\alpha_1-2^5\varepsilon)^2\alpha'}{R^2}\, .
\label{tadpole}
\eea

This divergence can be reproduced in the effective field theory limit.
For instance we consider the contribution from one of the $2^3$ states
present in (\ref{tadpole}) in four dimensions, descring the radion
$\varphi$ associated with the two  compact dimensions of radius $R$. Note
that in section 2 we considered fixed  the string coupling and thus
constant the ten dimensional dilaton. However,  once we turn on
fluctuations, it is easy to see that the fields that have  orthogonal
kinetic terms are the radion and the four-dimensional dilaton. It follows
that the relevant part of the action which describes the  fluctuations of
the radion is:
\bea
\int d^4x\,
\left[\, \frac{1}{2 \kappa^2}\, {\cal R}^{(4)}+
(\partial\varphi)^2+ e^{\displaystyle -\kappa \varphi}\frac{M_s^4}{g_s}V_0\,
\right]\, .
\label{act}
\eea
where the constant $V_0$ is given by the total tension of all branes and 
orientifolds. For the models we described in section 3.2, it is given by:  
\bea
V_0= (\alpha_1-2^5\varepsilon) \frac{(\alpha_1-2^5\varepsilon)}{2 (2\pi)^2 }
\label{V0}
\eea
Using the action (\ref{act}), one can compute the vacuum diagram that gives 
rise to the one-loop quadratic divergence (\ref{tadpole}); it is given by the 
massless on-shell propagator of the radion ending on two tadpoles:
\bea
\frac{M_s^8}{g_s^2}\frac{\kappa^2 V_0^2}{2 p^2}\bigg|_{p^2=0}=
\frac{1}{2^5\pi}\frac{M_s^4}{(2\pi)^4}
\frac{(\alpha_1-2^5\varepsilon)^2}{R^2} \frac{1}{p^2}
\label{qfttad}
\eea

To compare with the string theory side, we notice that:
\bea
\int dl= \int dl q^{\frac {p^2 \alpha'}{4}}\bigg|_{p^2=0}= \frac {2}
{\pi \alpha'p^2}\bigg|_{p^2=0} \, .
\eea 
It follows that the field theory result (\ref{qfttad}) reproduces the 
quadratic divergence of the string expression (\ref{tadpole}) (up to
the multiplicative numerical factor $2^3$ giving the massless states 
degeneracy). 

On the other hand, taking first the limit
$R\to\infty$, one finds additional contributions from the $\theta_4$
terms and one is left over with a logarithmic divergence $dl/l$, as
$l\to\infty$. This phenomenon is related to the fact that tadpoles do not
vanish locally (even when they vanish globally for $\alpha_1=32$
and $\varepsilon=+$), and thus, a new contribution arises in the
decompactification limit
\cite{ads1}.

A simple inspection shows that the logarithmic divergence comes entirely
from the M\"obius amplitude, proportional to
$\hat{\theta}_2^4/\hat{\eta}^{12}$ in eqs.~(\ref{GCKAM}) and
(\ref{GCKAMII}), as expected since it is the only source of supersymmetry
breaking in the $R\to\infty$ limit. This divergence is cut-off by the
size of the transverse dimension and leads in four dimensions ($d=4$) to
\bea
\Lambda_{UV}|_{I}&\simeq&\frac{1}{(4\pi^2\alpha')^{2}}
\int^{\infty}\frac{dl}{l}\,
2^4\times \frac{2^6}{2^8}(\alpha_1\varepsilon\theta_3^2-
\delta_2\theta_3\theta_4)(\frac{iR^2}{2\alpha'l})\, ,\\
\Lambda_{UV}|_{II}&\simeq&\frac{1}{(4\pi^2\alpha')^{2}}
\int^{\infty}\frac{dl}{l}\,
2^4\times \frac{2^6}{2^8}(\alpha_1\varepsilon\theta_3^2-
\beta_2\theta_4^2)(\frac{iR^2}{2\alpha'l})\, .
\eea
As a result, in the limit $R\to\infty$, we find a logarithmic behavior:
\bea
\Lambda_{UV}(R) &\simeq&  \alpha M_s^4\,\ln R\, ,
\label{uvdiv}
\eea
with
\bea
\alpha =\frac{8}{(4\pi^2)^{2}}\,(\alpha_1\varepsilon-\delta_2\,)
={\bigg|}_{\varepsilon=+}\ \frac{1}{\pi^4}
(n_{\pi \pi}+n_{\pi 0}+{\bar n}_{0 0}+{\bar n}_{0 \pi})
\label{alphaI}
\eea
for model $I$, and
\bea
\alpha =\frac{8}{(4\pi^2)^{2}}\,(\alpha_1\varepsilon-\beta_2\,)
={\bigg|}_{\varepsilon=+}\ \frac{1}{\pi^4}
({\bar n}_{\pi \pi}+n_{\pi 0}+{\bar n}_{0 0}+n_{0 \pi})
\label{alphaII}
\eea
for model $II$. Here, we gave the values only for the case
$\varepsilon=+$ which has no tree-level tadpoles (for $\alpha_1=32$).

Note that the coefficient $\alpha$ of the logarithm is in fact
proportional to the difference between the numbers of massless fermions
and bosons, in the decompactification limit. The remaining part of the
potential leads to a constant $\beta$, plus contributions  which vanish
in the large radius limit. In order to obtain the effective potential in
the  Einstein frame, we perform an appropriate rescaling as explained 
in Section 2, which introduces an overall factor $1/R^4$:
\bea
V^{(1)}_{eff}(R) =\frac{1}{R^4}V_1(R)
\stackrel{R \rightarrow \infty}{\simeq} 
\frac{1}{R^4}\left(\alpha\,\ln (RM_s)+\beta\right)\, ,
\label{pot}
\eea
where we used the notation introduced in Section 2.

Unfortunately, it is easy to see that for $\alpha$ positive, the
non-trivial extremum of the potential is either not consistent
with the approximation $R>>l_s$, or is a maximum. Indeed, the extremum
$R_0$ of $V_{eff}$ is given by:
\bea
R_0=l_s e^{{1\over4}-\frac{\beta}{\alpha}}\, ,
\eea
which is in the region $R>>l_s$ only for $\beta$ negative. However, since
the potential is positive asymprotically, this corresponds to a local
maximum, while the minimum corresponds to the run away value $R=\infty$.

Therefore, to find a minimum, we need a model where the coefficient
$\alpha$ of the logarithm is negative, implying that there is a surplus
of massless bosons versus massless fermions. In fact, as seen from
eqs.~(\ref{alphaI}) and (\ref{alphaII}), this is the case of the model
obtained using the choise
$\varepsilon=-$, for which:
\bea
\alpha{\big|}_{\varepsilon=-}\ =-\frac{1}{\pi^4}
(n_{00}+{\bar n}_{\pi \pi}+{\bar n}_{\pi 0}+n_{0 \pi})
\label{alphaIm}
\eea
in model $I$, and
\bea
\alpha{\big|}_{\varepsilon=-}\ =-\frac{1}{\pi^4}
(n_{00}+n_{\pi \pi}+{\bar n}_{\pi 0}+{\bar n}_{0 \pi})
\label{alphaIIm}
\eea
in model $II$. However, this model has non-vanishing tree-level dilaton
tadpole associated to a localized tree-level potential. In the weak
coupling limit $g_s<1$, this potential dominates over the constant
one-loop contribution $\beta$ of eq.~(\ref{pot}).\footnote{Note that in
the large radius limit, the one-loop quadratic UV divergence
(\ref{tadpole}) vanishes, consistently with the absence of tadpole
conditions in non-compact space. Of course, strictly speaking, this is
true only for more than two bulk dimensions so that the decoupling
decompactification limit exists. For $n=2$, one is left over with the
logarithmic correction (\ref{uvdiv}) that we take into account.} It can
be easily deduced from the one-loop divergence (\ref{tadpole}), and in
the string frame it reads (see eq.(\ref{V0})):
\bea
{1\over g_s}V_0={M_s^4\over 8\pi^2g_s}
(32+\alpha_1)=8{M_s^4\over\pi^2g_s}\, .
\label{vzero}
\eea

It follows that in this case $V_{eff}$ has a local minimum at the value:
\bea
R_0=l_s e^{{1\over 4}+\frac{ 8\pi^2}{g_s(n^++{\bar n}^+)}}\, ,
\eea
where $n^+$ (${\bar n}^+$) denotes the total number of D-branes (anti
D-branes) located on top of the anti-orientifold ${\bO}^+$-planes
(orientifold $\O^+$-planes), given in eqs.~(\ref{alphaIm}) and
(\ref{alphaIIm}): $n^+=n_{00}+n_{0\pi}$, 
${\bar n}^+={\bar n}_{\pi\pi}+{\bar n}_{\pi 0}$. By taking a value for
$g_s=g_{YM}^2\simeq .5\pm .1$, with $g_{YM}$ the gauge coupling at the
string scale, and varying $n^++{\bar n}^+$ between 1 and 32, it is very
easy to obtain hierarchical large values for the ratio $R_0/l_s$. In
Fig. 3, we plot the prediction for the string scale, as a function of the
gauge coupling for three different values of $n^++{\bar n}^+=3,4$ and 5.
\begin{figure}[htb]
\centering
\epsfxsize=5.0in
\epsfysize=3in
\epsfbox{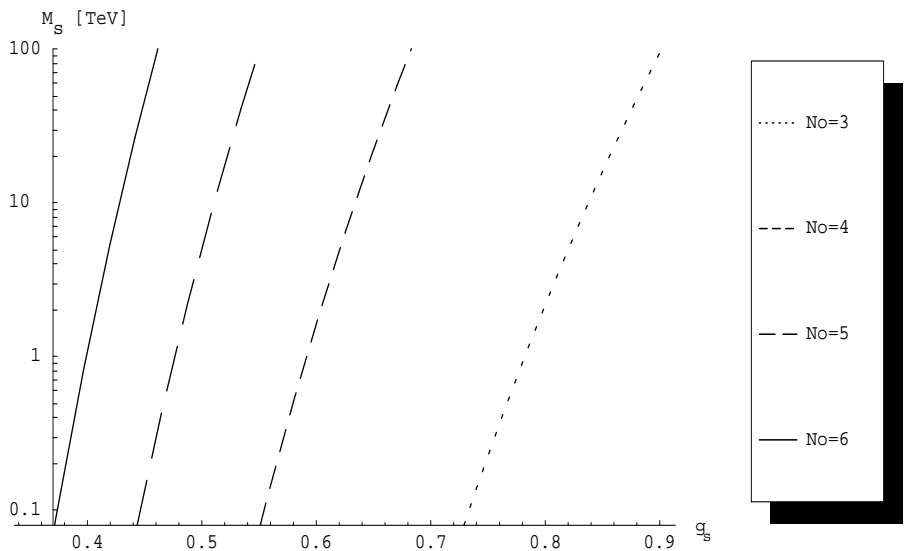}
\caption{ The string scale as a function of the coupling constant for different choices
for the number of non-supersymmetric branes. }
\end{figure}

\subsection{Generic case}

Here, we generalize the above results to models which have both types of
orientifold planes $\O^+$ and $\O^-$ (as well as $\bO^\pm$). Such
constructions can be obtained for instance by turning on (quantized)
antisymmetric tensor field background. The effective potential receives
a tree level contribution given by the total sum of tensions of
branes and orientifolds. In the case of orientifold 7-planes (toroidally
compactified in four dimensions), it reads:
\bea
{1\over g_s}V_0= \frac {M_s^4}{8\pi^2 g_s}{Q^{NS}}=
\frac{M_s^4} {8\pi^2 g_s}{(-8N^- - 8{\bar N}^-+8 N^+ +8{\bar N}^+ 
+n +{\bar n})}\, ,
\label{V0gen}
\eea
where $N^\pm$ (${\bar N}^\pm$) denote the number of $\O^\pm$ ($\bO^\pm$)
planes, while $n$ ($\bar n$) is the total number of D-branes
(anti-branes). 
On the other hand, in the large transverse radius limit, 
the one-loop contribution is dominated by the M\"obius 
amplitude, and thus, is proportional to the difference between the
numbers of massless fermionic and bosonic degrees of freedom.
In the case of two large extra dimensions, it leads to:
\bea
V_1(R) &\simeq& \frac {M_s^4}{\pi^4}({\cal N}_F-{\cal N}_B)\ln(RM_s)
\nonumber\\ &=&
\frac{M_s^4}{\pi^4} {(n^- + {\bar n}^- - n^+ -{\bar n}^+)}\ln(RM_s)
\label{V1gen}
\eea 
where ${\cal N}_B$ (${\cal N}_F$) is the number of massless bosons
(fermions) on the branes, and in analogy with $n^+$ (${\bar n}^+$),
$n^-$ (${\bar n}^-$) denotes the number of branes on anti-orientifolds 
$\bO^-$ (anti-branes on orientifolds $\O^-$). Here we neglected the
constant term, as it is suppressed by powers of the string coupling
$g_s$, compared to the tree-level contribution (\ref{V0gen}). 

Note that, by inspection of Table 1, the brane configurations appearing
in eq.~(\ref{V1gen}) are the only ones that break supersymmetry (at the
massless level), although, as explained in Section 3, there is still a
left-over non-linear supersymmetry. In all other configurations (branes
on $\O^\pm$, or anti-branes on $\bO^\pm$) supersymmetry is linearly
realized (at least at the massless level) and there are no logarithmic
corrections in the potential.

It follows that the potential has a minimum at:
\bea
R_0= e^{\frac{1}{4}+\frac{8N^+ +8{\bar N}^+ -8N^- - 8{\bar N}^- + n +
{\bar n}}{n^+ +{\bar n}^+ -n^- - {\bar n}^-}
\frac {\pi^2}{8 g_s}} l_s\, .
\eea
As in the previous subsection, it is easy to find examples where such a
formula gives a size for the compactification radius hierarchically
larger than the string length and in the desired range of values.

\section{One-loop contributions to the masses of bulk fields}
\subsection{Bulk scalar masses}

Let us consider a generic scalar $\Phi$ living in the closed string bulk.
Through its interactions with other particles, it gets a one-loop
correction to its mass $g^2\Sigma_\Phi \Phi^2$, where $g$ is the gauge
coupling constant, determined by the self energy at zero momentum:
\bea
\Sigma_\Phi(M_s, R) = \Sigma_{\parallel}(M_s, R) + 
\Sigma_{\perp}(M_s, R)\, ,
\eea
where $\Sigma_{\parallel}$ contains effects exclusively due to other
bulk particles, while $\Sigma_{\perp}$ comes from the presence
of the boundaries. Here, we restrict our discussion to the case
where the generated term quadratic in the field $\Phi$ is just a mass term, 
while in general there could be also terms localized on the boundaries 
of the form $\partial_I \Phi\partial^I \Phi$ or $M \Phi\partial_I \Phi$ 
with $I$ representing one of the transverse directions and $M$ a mass
scale.

We first consider the contributions from $\Sigma_{\parallel}(M_s, R)$.
In general,
this contribution can be extracted from the torus amplitude.
For the case of the radion field in the string models under study, 
assuming that the full potential has a minimum at a value $R_0$, 
one finds a mass of the order of 
\bea
\Sigma_{\parallel}\sim l_s^{2+n}/R_0^{4+n}\, ,
\label{sigmapar}
\eea
where $n$ is the number of large bulk dimensions with a common radius
$R_0$. Actually, the same result remains valid for any bulk scalar, which
follows simply from the behavior of the torus amplitude as
$1/R_0^{2+n/2}$, for large $R_0$.

The contribution $\Sigma_{\perp}$ originates from the other 
one-loop string surfaces with boundaries and crosscaps. As we have
pointed  out in section 2, and illustrated by the examples of section 3,
the radion field $\varphi$ acquires a potential given by (in the Einstein
frame):
\bea
R\to\infty:\quad V(R)\sim \left\{ \begin{array}{l}
 M_s^4 e^{-4\kappa \varphi}({\alpha \ln{(R/l_s)} + \beta})\qquad\quad {\rm for}\ \ n=2\\  
M_s^4 e^{-2n\kappa \varphi}({\alpha R^{(2-n)} +\beta}) \qquad\qquad 
{\rm for}\ \ n\neq 2 \end{array}\right.
\label{V2as}
\eea
where $\varphi= \kappa^{-1} \ln{(R/R_0)}$ as defined in section 2.

For the case of $n=2$, the potential has an extremum at 
$R_0$ with a squared mass $g^2\Sigma_\perp$
for the radion field $\varphi$, given by:
\bea
R_0=l_s e^{{1\over4}-\frac{\beta}{\alpha}}\qquad ; \qquad 
g^2 \Sigma_\perp = -32 \pi \alpha \frac{M_s^4}{M_P^2}
\label{r0}
\eea
so that a minimum at $R_0>>l_s$ requires $\alpha<0$, $\beta>0$ and
$-\beta/\alpha >>1$.

For $n\neq 2$, the potential in (\ref{V2as}) gives instead:
\bea
R_0=l_s \left\{-\frac{2n\beta}{(3n-2)\alpha}\right\}^{\frac{1}{2-n}}
\qquad ; \qquad 
g^2 \Sigma_\perp = 2n(2-n)\beta \frac{M_s^4}{M_P^2}\, .
\eea 
In order to have a minimum at $R_0$ with  $R_0>>l_s$ 
we need $ \beta <0$ and $-\beta/\alpha << 1$ for $n>2$ or
$ \beta >0$ and $-\beta/\alpha >> 1$ for $n=1$. Of course, in both cases,
unlike the case $n=2$, a large hierarchy cannot be obtained naturally,
since there are no exponentials as in (\ref{r0}).

The contribution to the mass from 
$\Sigma_{\perp}\sim {M_s^4/M_P^2}\sim 1/R_0^n$
dominates always over the contribution (\ref{sigmapar}) from
$\Sigma_{\parallel}\sim 1/R_0^{4+n}$ which can therefore be neglected.
Moreover, for $n>2$, the effect of mixing among KK excitations can also be
neglected, as the supersymmetry breaking induced mass is
suppressed compared to the KK masses by powers of
$l_s/R_0$ for $n>2$, and by a loop factor in the case of $n=2$
($1/R_0^{n/2}<1/R_0$). Thus, for $n\geq 2$ the radion mass is of the
order of $\frac{M_s^2}{M_P}$.

The case of one extra dimension needs a more careful treatment.
Because the boundary interactions do not conserve KK momenta,
the vacuum polarization diagram associated with
$\Sigma_{\perp}$ will connect different mass levels as shown in Fig. 4.
We denote by $g({n})$ the coupling of a KK state with momentum
${n}/R$ with the boundary states\footnote{In the case of states 
localized at brane intersections $g({n})= g {\sqrt{2
-\delta_{n,0}}}\delta ^{- n^2 l_s^2/R^2}$ \cite{ttu, abl}.} appearing in
the one-loop self-energy, and define the tree-level 5D propagator
projected on the boundary (at the origin):
\bea
\Pi_{5}= \sum_{n} \frac{1}{-{p}^2 +\frac{{n}^2}{R^2}}\, ,
\eea
where $p$ is the momentum along the directions parallel to the boundary.
We consider now the one-loop correction to the propagator in the 
large radius limit,
$R>>l_s$. Since in this limit the series in $\Pi_{5}$ is convergent, we
can replace $g({n}) = g{\sqrt{2
-\delta_{n,0}}}$ independent of ${n} (n\neq 0)$ as long as $n < Rl_s$.
The resulting propagator, obtained by the Dyson resummation shown in Figure 4,
is  then given by $(\Pi_{5}^{-1} -g^2\Sigma_{\perp})^{-1}$. Thus, the
one-loop  corrected mass of $\Phi$ KK excitations 
can be computed through the equation $\Pi_{5}^{-1}
-g^2\Sigma_{\perp}= 0$ for ${p}^2 =M^2$.
\vskip 1cm
\begin{figure}[htb]
\centering
\epsfxsize=5.0in
\epsfbox{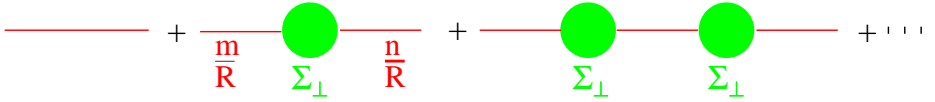}
\caption{The one-loop self-energy localized on the boundary introduces a 
mixing between the KK states and modifies the mass spectrum.  }
\end{figure}
\vskip 1cm
 
The mass eigenstates
are  then given by the solution of the equation:
\bea
\cot{(\pi R M)}= \frac{\pi R M}{\pi^2 R^2 g^2  \Sigma_\perp\, } .
\label{easy}
\eea
This equation can be solved numerically and leads to a tower
of states with masses:
\bea
M_n= \frac {n +\delta{n}}{R} , \, \qquad  n\geq 0 , \, \, \delta{n}<1\, ,
\eea
where our approximations remain valid for $n< RM_s$. 
The loop of $\Sigma_{\perp}$ leads to a potential for the projection
of $\Phi$ on the boundary which modifies the mass of all KK modes
by a shift $\delta{n}$. This shift can be approximated in the limiting cases 
of very small or very large one-loop self-energy compared to 
the tree-level masse. 

 First, in the absence of one-loop self-energy, i.e.
$\Sigma_{\perp} \rightarrow 0$, the equation (\ref{easy}) gives $\cot{(\pi
R M)}=\infty$ and thus $\delta{n} \rightarrow 0$. For the scalar fields considered here
 $R^2 g^2 \Sigma_{\perp} >> 1$,
we can approximate $\cot{(\pi R M)}\sim \frac{\pi}{2}(1-2\delta{n})$ 
and we obtain 
\bea
\delta{n} \sim \frac{1}{2} -\frac{2n+1}{2\pi R^2 g^2 \Sigma_\perp}\, ,
\eea
and thus $\delta{n} \rightarrow 1/2 $ which is the maximal possible
shift.  This means that the projection
of the field $\Phi$ on the boundary is vanishing in order to minimize the
potential $\sim g^2 \Sigma_\perp \Phi^2$.

For  the radion  case  in  the models under study in
this work,  we have observed that the main one-loop contribution comes
from the Klein, M\"obius strip and annulus. This is due to open (or
closed unoriented) strings and thus gives rise to a boundary potential.
Moreover, we have found above that for one dimension $\Sigma_{\perp}\sim
1/R$, and thus $R^2 g^2 \Sigma_{\perp} >> 1$, in the large radius limit.
This corresponds then to a shift $\delta{n} \rightarrow 1/2 $ and all bulk
 masses of each KK excitations remain of order $1/R$.

Finally, we would like to comment on the case of a small $\Sigma_{\perp}$ 
( which is not realized for the scalar fields 
discussed here)  leading to a small  shift $\delta n
<<1$ i.e. $R^2 g^2 \Sigma_{\perp} 
<< 1$. In this case we can approximate $\cot (\pi RM) \sim
1/(\pi\delta{n})$ which leads to:
\bea
(n+ \delta{n})\delta{n} \sim R^2 g^2  \Sigma_\perp\, .
\eea
The zero mode mass is then
\bea
M_0^2 = \frac{\delta{n}^2}{R^2} \sim  g^2  \Sigma_\perp\, ,
\eea
which is the familiar result obtained directly in the 
effective four-dimensional theory at scales below $1/R$, 
while the masses of the other KK modes are shifted by 
\bea
\delta{n}\sim \frac{ R^2 g^2  \Sigma_\perp\,}{n}\, ,
\eea
which decreases with $n$.

\subsection{The gravitino mass}

In this subsection, we compute the one-loop correction to the gravitino
mass and show that it is of the order of a loop factor times its
tree-level mass $1/2R$.
The one-loop induced gravitino mass can be read from 
the two-point correlation function of the corresponding vertex operators.
The gravitino vertex in type I string models, present at each boundary 
can be constructed from the type IIB R-NS and NS-R states by taking
the left-right symmetric linear combination.

The one-loop contribution to the gravitino mass can be splitted into two
parts, corresponding to the contributions of the $CP$-even and $CP$-odd
spin structures. For the computation, we will need the expression of the
vertex operators in the $(-^1\!\!/_2,0)$ ghost picture:
\bea
V_{(-^1\!\!/_2,0)}(k,\epsilon)\!\! &=&\!\! 
\epsilon_{\mu}\int[d^2z]:(e^{-\frac{\phi}{2}}(z)u_{\alpha}S^{\alpha}(z)
({\bar\partial}{X}^{\mu}+
\alpha'k\cdot\tilde{\psi}\tilde{\psi}^{\mu})({\bar z})\nonumber \\ 
&+& e^{-\frac{\tilde{\phi}}{2}}({\bar z})(\partial 
X^{\mu}+\alpha'k\cdot\psi\psi^{\mu})(z)\tilde{u}_{\alpha}\tilde{S}^{\alpha}({\bar 
z}))e^{ik\cdot X}(z,{\bar z}):
\eea
and in the $(-^1\!\!/_2,-1)$ ghost picture:
\bea
V_{(-^1\!\!/_2,-1)}(k,\epsilon)\!\! &=&
\!\!\epsilon_{\mu}\int[d^2z]:(e^{-\frac{\phi}{2}}(z)e^{-{\tilde{\phi}}}
({\bar z})u_{\alpha}S^{\alpha}(z)\tilde{\psi}^{\mu}
({\bar z})+c.c.)e^{ik\cdot X}(z,{\bar z}):\nonumber \\
\eea
where $X^\mu$ are the (non-compact) bosonic coordinates and $\psi^\mu$
($\tilde{\psi}^{\mu}$) their left (right) 2d fermionic superpartners. The
tilde stands for right-movers and $\phi$ is the super-reparametrization
ghost. $S^{\alpha}$ are the ten-dimensional spin fields, $u^{\alpha}$ the 
associated ten-dimensional spinors and $\epsilon^{\mu}$ the gravitino 
polarization vector. We will also need the worldsheet
supercurrent given by $T_F(z)=\psi\partial X(z) + \tilde{\psi} {\bar
\partial}{X}({\bar z})$. 

For the case of even spin structure, we need to compute:
\bea
A_{even} = <V_{(-^1\!\!/_2,0)}(k,\epsilon^1)
(z,{\bar z})V_{(-^1\!\!/_2,0)}(-k,\epsilon^2)(w,{\bar w})
:e^{\phi}T_F(v):>\, .
\label{even}
\eea
All terms in the vertex operators involving powers of the external
momentum $k$ are proportional, in the rest frame, to powers of $1/R$. The
only contraction that does not involve explicit powers of $k$ is
proportional to:
\bea
<{\bar \partial}{X}^{\mu}(z){\bar \partial}{X}^{\nu}
({\bar w})\partial X^{I}(v)e^{ik\cdot X}(z,{\bar z})
e^{-ik\cdot X}(w,{\bar w})>\, ,
\nonumber
\eea
plus permutations of left and right movers. Here, the index $I$ refers
either to (non-compact) spacetime or internal indices, collectively.
These terms contain three factors of $\partial X$ or ${\bar\partial}X$,
and thus, one of them will be contracted with an exponential or replaced
by an internal zero mode, leading to a power of $1/R$.

In the case of odd spin structure, we have to compute instead:
\bea
A_{odd}\! =<V_{(\!-^1\!\!/_2,-1)}(k,\epsilon^1\!)(z,{\bar z})
V_{(\!-^1\!\!/_2,0)}(\! -k,\epsilon^2)(w,{\bar w})e^{\phi}T_F(v)
e^{\tilde{\phi}}\tilde{T}_F({\bar v}):>\!.
\label{odd}
\eea
A reasoning similar to the above of even spin structure shows that the
only contractions that do not involve explicit powers of the external
momentum contain three factors of $\partial X$ or ${\bar\partial}X$,
which lead again to contributions proportional to $1/R$.

As a result, the one-loop induced mass term is of order $g_s/R$.
Note that as  for the case of scalar fields, bulk fermions masses might
receive one-loop contributions of two kinds: (i) from bulk interactions
(ii) from boundary interactions.
In the absence of tree level masses, the fermions do not receive
one-loop contribution from bulk interactions. This contribution is
then proportional to the splitting induced by the non-periodic
boundary conditions and vanishes in the decompactification limit.


\section{Summary}

In conclusion, in this work we studied the brane to bulk mediation
supersymmetry in type I string models with brane supersymmetry
breaking. Bulk scalar masses are generated at one-loop level and are of
order $M_s^2/M_P$, while bulk fermions acquire tree-level masses due to
the Scherck-Schwarz boundary conditions and are in general heavier, of
the order of the compactification scale $1/R$. Thus, when the string
scale is in the TeV region, the radion acquire a tiny mass and mediates
a new attractive universal force at micron distances. We computed its
coupling to matter and found that such a force could be detectable in
tabletop experiments that test gravity at short distances.

We also studied the particular case of two large bulk dimensions and
derived a general formula for the effective potential. We showed that its
minimization can stabilize the radion and fix the size of the bulk at
values that are hierarchically large than the string length, determining
the desired hierarchy between the Planck and string scales. It will be
interesting to apply this mechanism in model building of semi-realistic
string vacua.

\section*{Acknowledgements}

We would like to thank Ann Nelson and Riccardo Rattazzi for valuable 
comments on an earlier version of the paper.
This work was supported in part by the European Commission under RTN
contract HPRN-CT-2000-00148, and in part by the INTAS contract N 99-1-590. 
K.B. acknowledges the financial support provided through the European
Community's Human Potential Programme under contract HPRN-CT-2000-00131
Quantum Spacetime. A. L. thanks the Theory Division of CERN for its 
hospitality and partial financial support.

\end{document}